\documentclass[12pt]{iopart}
\usepackage{graphicx,color}
\usepackage[switch]{lineno}
\usepackage{bm}
\usepackage{bbm}
\usepackage{mathrsfs}
\def\Tr{\mbox{Tr}}

\begin{document}
\title{Transient quantum fluctuation theorems and generalized measurements}

\author{B. Prasanna Venkatesh$^1$,  Gentaro Watanabe$^{1,2}$ and Peter Talkner$^{1,3}$}
\address{$^1$Asia Pacific Center for Theoretical Physics (APCTP), San 31, Hyoja-dong, Nam-gu, Pohang, Gyeongbuk 790-784, Korea}
\address{$^2$Department of Physics, POSTECH, San 31, Hyoja-dong, Nam-gu, Pohang,
Gyeongbuk 790-784, Korea}
\address{$^3$Institut f\"{u}r Physik, Universit\"{a}t Augsburg, Universit\"{a}tsstra\ss e 1, D-86135 Augsburg, Germany}
\ead{balasubv@apctp.org}

\date{\today}
\begin{abstract}
The transient quantum fluctuation theorems of Crooks and Jarzynski  restrict and relate the statistics of work performed in forward and backward forcing protocols. So far, these theorems have been obtained under the assumption that the work is determined by projective energy measurements at the end and the beginning of each run of the protocols.
We found that one can replace these projective measurements only by special error-free generalized energy measurements with pairs of tailored, protocol-dependent post-measurement states that satisfy detailed balance-like relations. For other generalized measurements, the Crooks relation is typically not satisfied. For the validity of the Jarzynski equality, it is sufficient that the first energy measurements are error-free and the post-measurement states form a complete orthonormal set of elements in the Hilbert space of the considered system. Additionally, the effects of the second energy measurements must have unit trace. We illustrate our results by the example of a two-level system for different generalized measurements.        
\end{abstract}
\maketitle
\section{Introduction}\label{S1}
During the past one-and-a-half decades, the transient fluctuation theorems, known as Jarzynski equality \cite{J} and Crooks relation \cite{C} have attracted wide interest. They constitute exact relations for the non-linear response of systems in thermal equilibrium to arbitrary perturbations. As such they apply to a wide variety of systems, including for example bio-molecules, colloids, nuclear spins, cold atoms, and clusters of atoms, to name just a few. These theorems  have 
motivated numerous studies on the dynamics of small systems and also have revived and fertilized the field of nonequilibrium thermodynamics. Reviews of the different theoretical aspects  of transient fluctuation theorems applying to classical and quantum dynamics of open and closed systems can be found in \cite{JR,EHM,CHT,S}. 
Apart from theoretical studies, the fluctuation theorems have also inspired experimental studies \cite{Ciliberto} and have become the basis of a method for determining free energy differences of different configurations of large molecules \cite{BLR}.      

The central object of the transient fluctuation theorems is the {\it work} performed on a system by the variation of external parameters. Work together with heat constitute basic notions on which any thermodynamic description rests. Therefore, a convenient operational definition for work that is also applicable to quantum systems is of eminent importance in the thermodynamic description of biological and artificial devices on the nanoscale, the design of quantum engines, and quantum information processing.  

In the present work, we will restrict ourselves to the discussion of closed quantum systems. The validity of the fluctuation theorems is then based on the reversibility of Hamiltonian dynamics and on the assumption that the system is initially prepared in  canonical equilibrium described by a Gibbs state; out of this state,  
external time-dependent forces $\lambda(t)$ entering the system Hamiltonian $H(\lambda(t))$ drive the system away from  equilibrium according to a prescribed protocol  $\Lambda = \{\lambda(t)|0\leq t \leq \tau \}$ of finite duration $\tau$.
The work performed on a system in such a process is a statistical quantity, which can be characterized by a probability density function (pdf) $p_\Lambda(w)$.  

An associated backward process starts in thermal equilibrium 
at the initial inverse temperature $\beta$ and at the time-reversed final parameter values. The parameter values are then retraced according to the backward protocol $\bar{\Lambda} = \{\epsilon_\lambda \lambda(\tau -t)|0 \leq t \leq \tau \}$ where $\epsilon_\lambda$ is the parity of the parameter $\lambda$ under time-reversal, for example $\epsilon_\lambda=1$ for an electrical and $\epsilon_\lambda=-1$ for a magnetic field. The Crooks relation then 
connects the pdfs of the forward protocol $\Lambda$ and of the backward protocol $\bar{\Lambda}$ in the following way:  
\begin{equation}
p_\Lambda(w) = e^{-\beta(\Delta F -w)} p_{\bar{\Lambda}}(-w)\:,
\label{CR}
\end{equation}   
provided that the Hamiltonian satisfies the instantaneous time reversal symmetry $\theta^\dagger H(\lambda(t)) \theta = H(\epsilon_\lambda \lambda(t))$ where $\theta$ denotes the time-reversal operator.
In the Crooks relation, 
\begin{equation}
\Delta F = F(\lambda(\tau)) - F(\lambda(0))
\label{DF}
\end{equation}
 denotes the difference in free energies of canonical equilibrium systems at the force values  at the end and at the beginning of the protocol. The free energy 
\begin{equation}
F(\lambda) = - \beta^{-1} \ln Z(\lambda)
\label{F}
\end{equation}
is defined in terms of the respective canonical partition function 
\begin{equation}
Z(\lambda) = \Tr e^{-\beta H(\lambda)}\:.
\label{Z}
\end{equation}   
The Jarzynski equality 
\begin{equation}
\langle e^{-\beta w} \rangle = e^{-\beta \Delta F}
\label{JE}
\end{equation} 
follows as a direct consequence of the Crooks relation. 

For later use, we note that the Crooks relation can be expressed in terms of the characteristic function of work, defined as the Fourier transform of the work pdf, 
\begin{equation}
G_\Lambda(u) = \int dw e^{i u w} p_\Lambda(w)\:,
\label{Gw}
\end{equation} 
in the equivalent way, reading
\begin{equation}
Z(0) G_\Lambda(u) = Z(\tau) G_{\bar{\Lambda}}(-u + i \beta)\:.
\label{GG}
\end{equation}   

Other than in the classical case, direct experimental confirmations of the quantum transient fluctuation theorems are still missing. Huber \textit{et al}.\ \cite{HSDL} suggested an experimental proof by means of a cold atom sitting in a parabolic trap whose position can be varied. The difficulty of this, but also of other experiments testing transient quantum fluctuation theorems, lies in the measurement of the work performed on the system in a single realization. This requires two measurements of the system energy in the beginning and at the end of the protocol. In the theoretical investigations of the transient quantum fluctuation theorems \cite{K,Tas,TLH}, these measurements are projective \cite{vN}. That means that an energy eigenvalue $e_n$ with a corresponding eigenfunction $|n\rangle$ will be measured in a state described by a density matrix $\rho$ with the probability $p_n = \langle n|\rho | n\rangle$ and will leave the system immediately after the measurement in the state $|n\rangle \langle n |/p_n$.

One might suspect that a weaker, less invasive form of the energy measurements is simpler to realize in an experiment and still might leave the transient fluctuation theorems intact. This belief could be supported by the argument that any weaker, or less invasive measurement can be obtained as a projective measurement on a so-called ancilla which is typically an auxiliary two-level system that has to be brought into an entangled state with the actual system \cite{NC}. Under proper conditions, the measurement on the ancilla gives the required information on the system, possibly with somewhat reduced accuracy, while the back-action of the ancilla on the system is weaker than a direct measurement would impose.        

In two recent papers \cite{DCHFGV,MDP}, an experimental approach was suggested by means of which the characteristic function of work  is encoded in the reduced density matrix of an ancilla. Its realization requires a time-dependent system-ancilla coupling which follows the original force protocol amended by constant force periods of variable lengths. A tomography of the final ancilla state yields the characteristic function of work. In this way, projective energy measurements are avoided, yet, unlike the original setting, the forcing of the system is mediated by the coupling to the ancilla. The confirmation of the Crooks relation and the Jarzynski equality by means of this method was reported in \cite{BSMASOGDPS}. Another experiment utilizing circuit cavity QED was suggested in Ref.~\cite{CBKZH}. 

It though remains an open question whether generalized measurements can replace  the projective energy measurements and still conform with the fluctuation relations.
The paper is organized as follows. In Section~\ref{S2} we shortly review the theory of generalized measurements; for more extensive expositions we refer to the literature \cite{NC,WM}. In Section~\ref{S3} we derive the work distribution in the presence of generalized measurements and analyze in Section~\ref{S4} the restrictions on these measurements under which the Crooks relation and the Jarzynski equality hold. We illustrate our results by the example of a two level system whose Hamiltonian is suddenly changed, and end with some conclusions in Section~\ref{S5}. For the sake of the reader's convenience, some technical arguments are presented in Appendices \ref{A1}, \ref{A2}, and \ref{A3}.      
\section{Generalized measurements}\label{S2} 
A generalized measurement is formally characterized by a set of measurement operators $M_n$ satisfying the normalization condition 
\begin{equation}
\sum_n M^\dagger_n M_n =\mathbbm{1}\:,
\label{M1}
\end{equation} 
where $\mathbbm{1}$ denotes the identity operator on the Hilbert space of the considered system. 
In the present study, energy measurements are of central importance.
For that purpose, we need measurement operators $M_n$ that can identify the eigenstates $|n\rangle$ of a Hamiltonian $H$ with spectral representation 
\begin{eqnarray}
H&= \sum_n E_n \Pi_n\:,\\
\Pi_n &= |n\rangle \langle n|,
\label{SR} 
\end{eqnarray}
where, for the sake of notional simplicity, we restrict ourselves to non-degenerate energy-eigenvalues. The identification of an energy eigenstate 
by means of the measurement operator $M_n$ in general will not be perfect, rather the expression
\begin{equation} 
p(m|n) = \Tr M^\dagger_m M_m \Pi_n
\label{pmn}
\end{equation}
quantifies the likelihood of an erroneous assignment $m$ if a measurement is applied on the energy-eigenstate $|n\rangle$. 
For general energy measurements $M_n$, the post-measurement state is given by 
the density matrix
\begin{equation}
\rho_n = M_n \rho M^\dagger_n/p_n(\rho), 
\label{rhon}
\end{equation}
where $\rho$ denotes the density matrix immediately before the measurement and 
the expectation value of the ``effect'' $M^\dagger_n M_n$, 
\begin{equation}
p_n(\rho) =  \Tr M^\dagger_n M_n \rho, 
\label{pn}
\end{equation}
gives the probability with which the eigenstate $|n\rangle$ is observed. If a measurement is performed but its result is not registered, the post-measurement state is given by $\sum_n M_n\rho M^\dagger_n$. 

We want to emphasize that an energy measurement operator $M_n$ is primarily indicative of an energy eigenstate $|n\rangle$; the assignment of the energy eigenvalue occurs by the quantum number $n$. Therefore, the energy measurement operator $M_n$ need not depend on the actual energy eigenvalue $e_n$. Below we will make use of this possibility.

In the particular case when the conditional probability is sharp, 
$p(m|n) = \delta_{n,m}$, we call the measurement $M_n$ {\it error-free}.
Any error-free measurement operator $M_n$ is characterized by a single, normalized post-measurement state $|\psi_n\rangle$ in terms of which it can be written as
\begin{equation}
M_n = |\psi_n \rangle \langle n|, \quad \langle \psi_n|\psi_n \rangle =1\:.
\label{Mef}
\end{equation}
See Appendix \ref{A1} for details. In general, for different $n$, the post-measurement states need not be mutually orthogonal or linearly independent.

A special case of an error-free measurement is a projective measurement, for which $|\psi_n\rangle = |n \rangle$ and hence, $M_n = \Pi_n$. Projective measurements are the only measurements that are error-free and that reproduce the result of the first measurement in an immediate repetition of the same measurement. Error-free measurements which are not necessarily repeatable were introduced by Landau and Lifshitz in the first Russian edition of their Quantum Mechanics in 1948; for an English translation see \cite{LL}. According to the classification by Wiseman and Milburn \cite{WM}, error-free measurements are called sharp measurements. 
\section{Work statistics from generalized measurements}\label{S3} 
As already mentioned, to obtain the work which is performed on a system during a protocol $\Lambda$,  two energy measurements are required, one at the beginning, and the second one at the end of the protocol. Since, in general, the respective Hamiltonian operators $H(\lambda(0))$ and $H(\lambda(\tau))$ at the beginning and at the end will be different, their instantaneous sets of eigenfunctions $\{|n;0 \rangle \}$ and $\{|n;\tau \rangle \}$ and eigenvalues $\{e_n(0)\}$ and $\{e_n(\tau)\}$ will in general also differ.   
Therefore, different sets of measurement operators $\{M_n(0) \}$ and $\{M_n(\tau)\}$ will be needed to determine the work. The probability to register an energy eigenstate $|n;0\rangle$ of the Hamiltonian $H(\lambda(0))$ is given by (\ref{pn}) with 
\begin{equation}
\rho =\rho(0)\equiv Z^{-1}(\lambda(0)) e^{-\beta H(\lambda(0))}
\label{rho0}
\end{equation}
 describing the thermal equilibrium at the beginning of the protocol. The conditional probability to measure the eigenstate $|m;\tau\rangle$ of $H(\lambda(\tau))$ upon the observation of the eigenstate $|n;0\rangle$ then becomes 
\begin{equation}
p_\Lambda(m|n) = \Tr M^\dagger_m(\tau) M_m(\tau) U(\Lambda) M_n(0) \rho(0) M_n^\dagger(0)U^\dagger(\Lambda)/p_n(\rho(0)), 
\label{pL}
\end{equation}
where  $U(\Lambda)= U_{\tau,0}$ denotes the unitary time evolution for the period of the protocol, given by the solution of the Schr\"odinger equation 
\begin{equation}
i \hbar \partial_t U_{t,0} = H(\lambda(t)) U_{t,0}\:,\quad  \;U_{0,0} = \mathbbm{1} \:. 
\label{SE}
\end{equation}
With the resulting joint probability 
\begin{equation}
p_\Lambda(m,n)= \Tr U^\dagger(\Lambda) M^\dagger_m(\tau) M_m(\tau) U(\Lambda) M_n(0) \rho(0) M^\dagger_n(0) 
\label{pLmn}
\end{equation}
to measure first $n$ and then $m$, the pdf of work takes the form:
\begin{equation}
p_{\Lambda}(w) = \sum_{n,m} \delta(w - e_m(\tau)+e_n(0)) p_\Lambda(m,n)\:.
\label{pLw}
\end{equation} 
The cha\-rac\-ter\-istic function then takes the form
\begin{equation}
G_\Lambda(u) = Z^{-1}(0)\Tr U^\dagger(\Lambda) Q(u,\tau) U(\Lambda) R(u,0)\:,
\label{G}
\end{equation}
where
\begin{eqnarray}
\label{Q}
Q(u,\varsigma) &= \sum_m e^{i u e_m(\varsigma)} M^\dagger_m(\varsigma) M_m(\varsigma)\:,\\    
R(u,\varsigma) &= \sum_n e^{-iu e_n(\varsigma)} M_n(\varsigma) e^{-\beta H(\lambda(\varsigma))} M^\dagger_n(\varsigma)\:,
\label{R}
\end{eqnarray}
with $\varsigma = 0$ and $\tau$. Requiring  that the Crooks relation (\ref{GG}) also holds in the presence of generalized measurements, we find as a condition on the measurement operators the equation 
\begin{equation}
\fl
\Tr U^\dagger(\Lambda) Q(u,\tau) U(\Lambda) R(u,0)  
= \Tr U^\dagger(\Lambda) R(-u+i \beta,\tau) U(\Lambda) Q(-u+i\beta,0)\:.
\label{CC}   
\end{equation}
To obtain the right hand side, we expressed the time evolution of the reversed protocol, $U(\bar{\Lambda})$ in terms of that of the forward process by means of time reversal symmetry, i.e., by $U(\bar{\Lambda}) = \theta^\dagger U^\dagger(\Lambda) \theta $, \cite{CHT,AG}. 
The condition (\ref{CC}) is necessary and sufficient for the Crooks relation to be satisfied. It presents a central formal result of our paper.  

\section{Universal measurements}\label{S4} 
We ask now whether 
energy measurement operators exist that do not depend on details of the protocol but for which the Crooks relation is satisfied. More precisely, we want to identify {\it universal} measurements operators $M_n(\varsigma)$, $\varsigma = 0$ and $\tau$, that satisfy the Crooks relation
 for all possible protocols having the same pairs of initial and 
final energy eigenstates $\{|n;\varsigma\rangle\}$, 
independent of the particular values of the eigenenergies at the beginning and the end.
As a consequence, universal measurements operators must be independent of the temperature of the initial state and of the particular protocol. 
Obviously, projective measurements are universal, but are there also others?

\subsection{Crooks relation}\label{S41}
The answer to this question can be found by analyzing (\ref{CC}). As just mentioned, for universal measurement operators,  (\ref{CC}) must be satisfied for any protocol that connects the initial and the final Hamiltonian within a finite time $\tau$. Therefore, it must also hold for the following particular protocol consisting of two sudden switches: One at the beginning and the second one at the end of the protocol. The first switch suddenly changes the Hamiltonian from its initial form $H(\lambda(0))$ to $G \eta/\tau$. Here, $\eta$ is a positive time scale and, as a Hamiltonian, $G$ must be hermitean, but otherwise it is arbitrary. 
The second switch results in the final Hamiltonian $H(\lambda(\tau))$. For this protocol, the time-evolution from $t=0$ to $t=\tau$ is given by $U(\Lambda) = e^{-i G \eta/\hbar}$ because the initial and final moments of the protocol at which the Hamiltonian differs from $G$ do not contribute to $U(\Lambda)$. With $G$ being an arbitrary Hamiltonian and $\eta$ an arbitrary time,  
the condition (\ref{CC}) must be satisfied for any unitary operator $U(\Lambda)$ in the case of universal measurements. As the simplest choice, one may take $\eta =0$ leading with $U(\Lambda) =\mathbbm{1}$ to 
\begin{equation}
\Tr Q(u,\tau) R(u,0) = \Tr R(-u+i\beta; \tau) Q(-u+i\beta;0)\:.
\end{equation} 
Using the explicit expressions (\ref{Q}) and (\ref{R}), we find 
\begin{equation}
\sum_{m,n} e^{i u (e_m(\tau) - e_n(0))} e^{-\beta e_n(0)} C_{m,n} =0\:,
\label{C}
\end{equation}
 where 
\begin{eqnarray}
C_{m,n}= \sum_k \left[e^{\beta(e_n(0) - e_k(0) )} A_{m,n,k} - e^{\beta(e_m(\tau)-e_k(\tau))} B_{m,n,k} \right]\:,\label{CAB}
\end{eqnarray}
with
\begin{eqnarray}
A_{m,n,k}&= \Tr M^\dagger_m(\tau)M_m(\tau) M_n(0)\Pi_k(0) M^\dagger_n(0)\:, \nonumber\\
B_{m,n,k}&=  \Tr M_m(\tau) \Pi_k(\tau) M^\dagger_m(\tau) M^\dagger_n(0)  M_n(0)\:.
\label{AB}
\end{eqnarray}
Because  the coefficients $A_{m,n,k}$ and $B_{m,n,k}$ are independent of $u$, $\beta$, $e_m(0)$, and $e_n(\tau)$ for universal measurements, the terms $C_{m,n}$ must vanish individually, further implying that $A_{m,n,k} =0$ for $k\neq n$ and $B_{m,n,k}=0$ for $k \neq m$. Taking the sum of $A_{m,n,k}$ over all $m$ and, accordingly, the sum of $B_{m,n,k}$ over all $n$ one obtains for the conditional probabilities 
\begin{equation}
p_\varsigma(m|k) = \Tr  M_m(\varsigma)\Pi_k(\varsigma) M^\dagger_m(\varsigma) = \delta_{m,k}\:.
\label{psig}
\end{equation}
Hence, both the first ($\varsigma =0$) and the second ($\varsigma = \tau$) energy measurements must be error-free. Using the expression (\ref{Mef}) for error-free measurements, we find as a necessary condition that the measurement operators assume the form
\begin{equation}
M_n(\varsigma) = |\psi_n(\varsigma)\rangle \langle n;\varsigma|\:,\quad \varsigma =0,\tau \:,
\label{Mnef}
\end{equation}
where $|n;\varsigma\rangle$ denotes the $n$th instantaneous eigenstate of $H(\lambda(\varsigma))$ and $|\psi_n(\varsigma) \rangle$ the respective, normalized post-measurement state.
Putting the explicit form of the measurement operators (\ref{Mnef}) into the necessary and sufficient condition (\ref{CC}) for the validity of the Crooks relation, one obtains 
\begin{equation}
\fl
\sum_{n,m} e^{i u(e_m(\tau)-e_n(0))} e^{-\beta e_n(0)} \left [ |\langle \psi_n(0) |U^\dagger(\Lambda) | m; \tau \rangle|^2 
 - |\langle n;0 |U^\dagger(\Lambda) |\psi_m(\tau) \rangle|^2 \right ] =0\:.
\end{equation}
For a universal measurement this equation must hold for all values of $u$ and $\beta$ and hence the square brackets must vanish for all values of the indices $n$ and $m$, i.e.,  
\begin{equation}
|\langle m; \tau |U(\Lambda)| \psi_n(0) \rangle |^2 = |\langle\psi_m(\tau) |U(\Lambda)| n; 0\rangle|^2\:.
 \label{db}
\end{equation}
As explained above, for a universal measurement, this condition must also be satisfied for all possible unitary maps $U(\Lambda)$. This requirement leads after some algebra, for details see Appendix ~\ref{A2}, to the conclusion that the post-measurement states $|\psi_m(\varsigma) \rangle$ have to coincide with the eigenstates $|m;\varsigma\rangle$, for $\varsigma=0$ and $\tau$, 
possibly up to irrelevant phase factors.
In other words, our main conclusion is that projective measurements are the only universal measurements that lead to the Crooks relation.

However, one may design the post-measurement states such that they satisfy the relation (\ref{db}) for a given protocol. For this particular, non-projective error-free measurement, the Crooks relation and consequently also the Jarzynski equality are satisfied. It is interesting to note that (\ref{db}) is analogous to the generalized detailed balance condition for transitions in forward and backward force protocols  when the respective initial states are microcanonical \cite{TMYH}. 
In the present case of generalized measurements, this generalized detailed balance condition requires that the respective transition probabilities from the post-measurement states of the first to the target states of the second measurement are the same for the forward and the backward protocol, i.e.,
\begin{equation}
\mbox{Prob}(|\psi_n(0) \rangle \stackrel{\Lambda}{\to} |m; \tau \rangle) = \mbox{Prob}(\theta |\psi_m(\tau) \rangle \stackrel{\bar{\Lambda}}{\to} \theta |n; 0 \rangle)\:.
\end{equation} 
This follows from the fact that the right hand side of (\ref{db}) can equally be expressed in terms of the time-reversed dynamics $U(\bar{\Lambda})$ as $|\langle \psi_m(\tau)|U(\Lambda) |n;0\rangle|^2 = |\langle n;0|\theta^\dagger U(\bar{\Lambda})\theta |\psi_m(\tau)\rangle|^2$.
\subsection{Jarzynski equality}\label{S42}
Requiring only the validity of the Jarzynski equality for universal generalized measurements, one finds 
that the initial energy measurement must be error-free,
\begin{equation}
M_n(0) = |\psi_n(0) \rangle \langle n;0| 
\end{equation}
with post-measurement states $\{ |\psi_n(0) \rangle \} $ that form a complete 
orthonormal basis set, hence satisfying
 \begin{eqnarray}
\langle \psi_n(0)|\psi_k(0) \rangle &= \delta_{n,k}\:,\\
\sum_n |\psi_n(0)\rangle \langle \psi_n(0) | &= \mathbbm{1} \:. 
\label{comp}
\end{eqnarray}
Moreover, the second energy measurement must be determined by effects  $M^\dagger_m(\tau)M_m(\tau)$ that have unit trace, $\Tr M^\dagger_m(\tau)M_m(\tau)=1$.
For a proof of these necessary and sufficient conditions see Appendix \ref{A3}.

\section{Examples}\label{S5}  
We illustrate our results by the example of a two level system which undergoes a sudden switch from 
\begin{equation}
H(0) = \epsilon \sigma_z/2
\label{H0}
\end{equation}
to 
\begin{equation}
H(\tau=0^+) =\epsilon \sigma_z/2+ \Delta \sigma_x\:,
\label{H1}
\end{equation} 
where $\sigma_x$ and $\sigma_z$ denote Pauli spin matrices. The eigenvectors and eigenvalues of $H(0)$ are given by 
\begin{eqnarray}
|1;0\rangle & = |-\rangle,\quad  |2;0\rangle = |+\rangle \\ 
e_{1,2}(0) &= \mp \epsilon/2
\label{ev0} 
\end{eqnarray} 
and those of $H(\tau)$
by 
\begin{eqnarray}
|1; \tau\rangle &= \sin(\gamma) |+\rangle - \cos(\gamma) |-\rangle,\quad 
|2; \tau\rangle = \cos(\gamma) |+\rangle + \sin(\gamma) |-\rangle\:, \\ e_{1,2}(\tau) &=\mp \sqrt{\epsilon^2/4 +\Delta^2}\:,
\label{ev1}
\end{eqnarray} 
where in the energy expressions the upper (lower) sign corresponds to the label 1 (2), and $|\pm \rangle$ denote the eigenstates of $\sigma_z$ with the respective eigenvalues $\pm 1$. The angle $\gamma$ follows from $\tan (2 \gamma) =2 \Delta /\epsilon$. In the following examples we set $\Delta = \epsilon/2$. 

\begin{figure}
\includegraphics[width=0.6\textwidth]{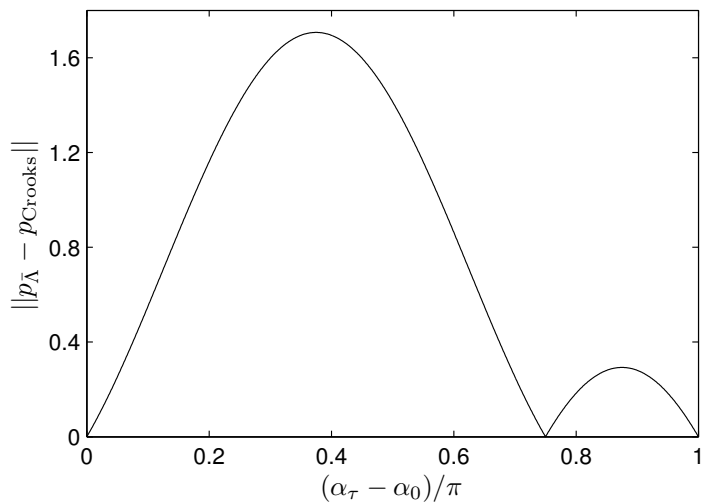}
\caption{The distance between the work probabilities of the backward protocol and the prediction of the Crooks relation for a two-level atom undergoing a sudden switch. The energy measurements are error-free with post-mesurement states that are characterized by mixing angles $\alpha_0$ and $\alpha_\tau$. For $\alpha_\tau - \alpha_0=0$, $3 \pi/4$, and $\pi$ the generalized detailed balance relation (\ref{db}) is satisfied, and therefore the Crooks relation holds, even though these measurements are not projective.}
\label{f1}
\end{figure}

We first consider error-free measurements of the form $M_n(\varsigma) = |\psi_n(\varsigma) \rangle \langle n;\varsigma|$ with $|n,\varsigma\rangle$ as defined by the eigenstates of the Hamiltonians $H(\varsigma)$ with $\varsigma = 0,\tau$ and the orthogonal post-measurement states
\begin{eqnarray} 
|\psi_1(\varsigma)\rangle &= \cos(\alpha_\varsigma)|1;\varsigma \rangle -\sin(\alpha_\varsigma) |2; \varsigma \rangle\:, \nonumber\\
|\psi_2(\varsigma)\rangle &= \sin(\alpha_\varsigma)|1;\varsigma \rangle +\cos(\alpha_\varsigma) |2; \varsigma \rangle\:.
\label{pm1}
\end{eqnarray}
In Figure \ref{f1}, we compare the actual probability distribution of the four possible outcomes of the work for the backward process $p_{\bar{\Lambda}}(w)$ with that of the the prediction $p_{\mathrm{Crooks}}(w) =e^{\beta(\Delta F + w)} p_\Lambda(-w)$ resulting from the Crooks relation (\ref{CR}) based on the probability distribution for the forward process. As a quantitative measure we used the sum of the absolute differences of these,
 $||p_{\bar{\Lambda}} -p_{\mathrm{Crooks}} || = \sum_{i=1}^4 | p_{\bar{\Lambda}}(w_i) -p_{\mathrm{Crooks}}(w_i)|$.
This distance may take values between $0$ and $2$, where $0$ indicates perfect agreement. The maximal distance $2$ reveals two completely disjoint distributions. In this example the initial and the final energy measurements are both error-free with orthogonal post-measurement states. Since the effects of the second measurement have unit trace, $\Tr M^\dagger_m(\tau)M_m(\tau) =1$, the Jarzynski equality is satisfied.  

\begin{figure} 
\includegraphics[width=0.6\textwidth]{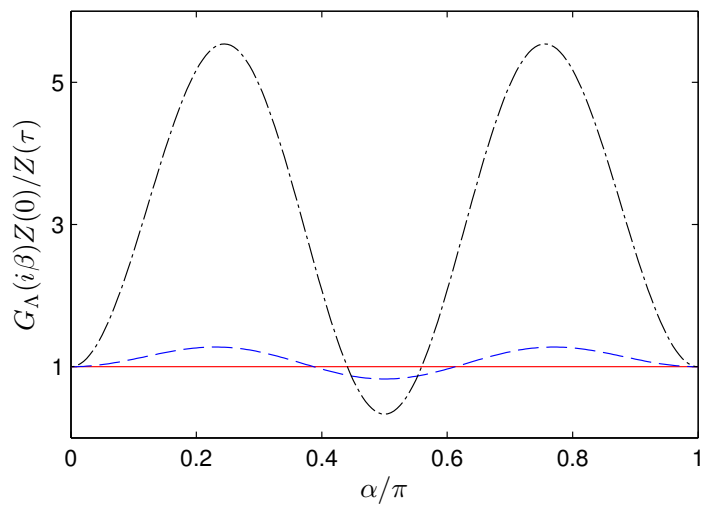}
\caption{(Color online) Deviations from the Jarzynski equality quantified by the ratio $G_\Lambda(i\beta) Z (0)/Z (\tau)$ are displayed for the measurement operators (\ref{M2}) as a function of the mixing angle $\alpha$ where $\alpha =0$ and $\pi$ correspond to projective measurements with correct assignment, $M_n=\pm \Pi_n$ respectively. For $\alpha =\pi/2$, $M_n=\Pi_{3-n}$. 
Deviations of the ratio from unity decrease with increasing temperature;
$\beta\epsilon=3$ (dash-dotted line), $1$ (dashed line),
and $4 \times 10^{-6}$ (solid line).
}
\label{f2}
\end{figure}

In the second example, illustrated in Figure \ref{f2}, we consider the same protocol but now choose a linear combination of projective energy measurements given by  
\begin{eqnarray}
M_1(\varsigma)&= \cos(\alpha) \Pi_1(\varsigma) + \sin(\alpha) \Pi_2(\varsigma)\:,\nonumber\\
M_2(\varsigma)& = \sin(\alpha) \Pi_1(\varsigma) + \cos(\alpha) \Pi_2(\varsigma)
\label{M2}
\end{eqnarray} 
with $\Pi_n(\varsigma) = |n;\varsigma \rangle \langle n;\varsigma|$. These measurement operators give successful and wrong assignments with probabilities $p(1|1) =p(2|2) = \cos^2 (\alpha)$ and $p(1|2)=p(2|1) = \sin^2 (\alpha)$, respectively.  The deviation from the Jarzynski equality can be quantified by the product
\begin{eqnarray}
\fl
\frac{Z (0)}{Z (\tau)} G_\Lambda(i \beta) &= p^{\mathrm{eq}}_1(\tau) \left [\cos^2(\alpha) \sin^2(\alpha) e^{\beta \epsilon}  
+\cos^2(\alpha)  
+ \sin^4(\alpha) e^{-\beta \epsilon}  \right] \nonumber\\
&\quad+ p^{\mathrm{eq}}_2(\tau) \left [\sin^4(\alpha) e^{\beta \epsilon} 
+\cos^2(\alpha) 
+ \cos^2(\alpha) \sin^2(\alpha) e^{-\beta \epsilon} \right ]\:,
\label{Ja}
\end{eqnarray}
where $p^{\mathrm{eq}}_n(\tau) =e^{-\beta e_n(\tau)}/(e^{-\beta e_1(\tau)}+e^{-\beta e_2(\tau)} )$. A value of unity indicates the validity of the Jarzynski equality and deviations quantify its failure. For the angles $\alpha = 0$ and $\pi$, the measurement is projective and hence the Jarzynski equality is satisfied. At the mixing angle $\alpha = \pi/2$, the measurement is also projective, but the assignment of energies is inverted. 
The deviations are more pronounced at low temperatures and diminish with increasing temperature.
\\
In the second example the effects of the second measurements have unit trace, $\Tr M^\dagger_1(\tau) M_1(\tau) =\Tr M^\dagger_2(\tau) M_2(\tau) =1$ whereas the first energy measurements are not error-free, and hence the initial energy measurements cause the violation from the Jarzynski equality. 

In the third example we consider error-free measurements with complete post measurement states initially but with final energy measurements whose effects have traces which may differ from unity. 
Choosing  the second measurements as
\begin{eqnarray}
M_1(\tau) &= \Pi_1(\tau) + \cos(\vartheta) \Pi_2(\tau) \nonumber\\
M_2(\tau) &= \sin(\vartheta) \Pi_2(\tau)
\label{sdm}
\end{eqnarray}
we find for the deviation from the Jarzynski equality
\begin{equation}
\frac{Z(0)}{Z(\tau)} G_\Lambda(ib) = 1+ (p_1^{\mbox{\footnotesize{eq}}}(\tau) -p_2^{\mbox{\footnotesize{ eq}}}(\tau))\cos^2 (\vartheta)\:, 
\label{ZZG}
\end{equation}    
Only at $\vartheta=\pi/2$ both effect traces $\Tr M^\dagger_1 M_1 = 1+ \cos^2 (\vartheta)$, $\Tr M^\dagger_2 M_2 = \sin^2 (\vartheta)$ become unity and consequently the Jarzynski equality holds only for this particular angle. The deviations from the Jarzynski equality again are largest for low temperatures for which the population difference $|p^{\mbox{\footnotesize{eq}}}_1(\tau)-p^{\mbox{\footnotesize{eq}}}_2(\tau)|$ approaches unity, and goes to zero in the limit of infinite temperature.  

\section{Conclusions} \label{S6}
We found that generalized energy measurements at the beginning and the end of a force protocol lead to work pdfs which in general violate the transient fluctuation theorem of Crooks. 
The Jarzynski equality holds for universal energy measurements if the first one is error-free with a complete set of orthonormal post-measurement states, and the second one has effects with unit trace. The Crooks relation remains valid for other than projective measurements only if both the initial and final energy measurements are error free with post-measurement states that are adapted to the actual protocol in such a way that the transition probabilities satisfy the generalized detailed balance relation (\ref{db}). Whether such measurement operators can be experimentally implemented in an easier manner than projective energy measurements remains an important open question.

\ack
We acknowledge support by the Max Planck Society, the Korea Ministry of Education,
Science and Technology (MEST), Gyeongsangbuk-Do, Pohang City, 
as well as by the Basic Science
Research Program through the National Research Foundation of Korea
funded by the MEST (No. 2012R1A1A2008028).
\appendix
\renewcommand\thesection{\Alph{section}}
\section{Error-free measurements}
\label{A1}
Any energy measurement operator may be represented by means of energy eigenfunctions $|k \rangle$ as
\begin{equation}
M_m = \sum_{k,l} g^m_{k,l} |k\rangle \langle l|\:,
\label{Mn}
\end{equation}
and accordingly the adjoint operator $M^\dagger_m$ as
\begin{equation}
M^\dagger_m =\sum_{k,l} (g^m_{k,l} )^* |l\rangle \langle k|\:,
\label{Mna}
\end{equation}
where the coefficients $g^m_{k,l}$ are complex numbers which, as a consequence of (\ref{M1}), fulfill the normalization condition 
\begin{equation}
\sum_{m,k}(g^m_{k,l'})^* g^m_{k,l} = \delta_{l,l'}\:.
\label{gnorm}
\end{equation}
The conditional probability $p(m|n)$, defined in (\ref{pmn}), specifying the reliability of the measurement can be expressed in terms of the coefficients $g^m_{k.l}$ 
\begin{equation}
p(m|n) = \sum_k |g^m_{k,n}|^2\:.
\label{pg}
\end{equation}
For error-free measurements satisfying $p(n|m) =\delta_{m,n}$, this implies
$g^m_{k,n}=0$ for $n \neq m$, and hence
\begin{equation}
M_m=|\psi_m \rangle \langle m|\:,
\label{Mnef2}
\end{equation}
where 
\begin{equation}
|\psi_m \rangle = \sum_k g^m_{k,m} |k \rangle .
\label{psigm}
\end{equation}
The normalization $\langle \psi_n|\psi_n \rangle =1$ follows immediately from the normalization condition (\ref{M1}). This completes the proof of the expression (\ref{Mef}) for error-free measurement operators.

Finally we prove that the post-measurement states $|\psi_n\rangle$ of an error-free measurement provide a complete orthonormal basis-set if they allow a partition of the unity, i.e. if
\begin{equation}
\sum_n |\psi_n\rangle \langle \psi_n | = \mathbbm{1} 
\label{pu}
\end{equation}   
holds.
Multiplying both sides by $\langle \psi_k|$ from the left and $|\psi_k \rangle$
from the right we obtain
\begin{equation}
\langle \psi_k | \psi_k \rangle^2 + \sum_{n (\neq k)} |\langle \psi_k | \psi_n \rangle|^2 = \langle \psi_k | \psi_k \rangle
\end{equation}
With the normalization $\langle \psi_k | \psi_k \rangle =1$ we immediately obtain the missing property of orthogonality  $\langle \psi_k | \psi_n \rangle = 0$ for $k\neq n$.

\section{Crooks relation}\label{A2}
Denoting 
\begin{eqnarray}
\Pi_n(\varsigma)&=|n;\varsigma\rangle \langle n;\varsigma| \label{Pin}\\
\Sigma_n(\varsigma)&= |\psi_n(\varsigma) \rangle \langle \psi_n(\varsigma) |\:,\label{Sin}
\end{eqnarray}
we may write ~(\ref{db}) as
\begin{equation}
\Tr \:\Pi_m(\tau) U(\Lambda) \Sigma_n(0) U^\dagger(\Lambda) = \Tr \:\Sigma_m(\tau) U(\Lambda) \Pi_n(0) U^\dagger(\Lambda)\:.
\end{equation}
With the spectral representation of the unitary operator 
\begin{equation}
U(\Lambda) = \sum_k e^{i \phi_k} |\varphi_k \rangle \langle \varphi_k |
\end{equation}
this can be further transformed into
\begin{eqnarray}
%\fl
&
\fl
\sum_{k,l} e^{-i(\phi_k-\phi_l)} \left [ \langle \varphi_k| \Pi_m(\tau) | \varphi_l \rangle \langle \varphi_l | \Sigma_n(0) |\varphi_k \rangle
-\langle \varphi_k| \Sigma_m(\tau) | \varphi_l \rangle \langle \varphi_l | \Pi_n(0) |\varphi_k \rangle \right ] \nonumber \\
&\qquad= 0\:.
\label{c1}
\end{eqnarray}
Because for universal measurements the time evolution operator $U(\Lambda)$ may be considered as an arbitrary unitary operator (see the discussion of the first paragraph of Section~\ref{S41}), the phases $\phi_k$ as well as the complete orthonormal basis set $\left \{ |\varphi_k \rangle\right \} $ are arbitrary. Therefore, the brackets under the sum of (\ref{c1}) must vanish individually. Rearranging the terms we obtain
\begin{equation}
\frac{\langle \varphi_k|\Pi_m(\tau)|\varphi_l\rangle}{\langle \varphi_k|\Sigma_m(\tau)|\varphi_l\rangle} = \frac{\langle \varphi_k|\Pi_n(0)|\varphi_l\rangle}{\langle \varphi_k|\Sigma_n(0)|\varphi_l\rangle}\equiv c_{k,l}\:.
\label{ckl}
\end{equation}
Hence, the ratios of matrix elements do neither depend on $n$ nor on $m$.
Using again the notation $\varsigma =0$ and $\tau $ we may treat both ratios simultaneously.  They can be rewritten with the help of (\ref{Pin}) and (\ref{Sin}) as
\begin{equation}
\frac{ \langle \varphi_k|m;\varsigma\rangle}{\langle \varphi_k|\psi_m(\varsigma)\rangle} = c_{k,l} \frac{\langle \psi_m(\varsigma) |\varphi_l\rangle}{\langle m;\varsigma |\varphi_l\rangle}\:.
\label{kkll}
\end{equation}
Obviously, the constant $c_{k,l}$ is given by a product of the form $(d_k d^*_l)^{-1}$ with
\begin{equation}
\frac{ \langle \varphi_k|m;\varsigma\rangle}{\langle \varphi_k|\psi_m(\varsigma)\rangle} = (d_k)^{-1}
\label{dk}
\end{equation}
Solving for ${\langle \varphi_k|\psi_m(\varsigma)\rangle}$, multiplying with $|\varphi_k\rangle$, and finally summing over all $k$ yields
\begin{equation}
|\psi_m(\varsigma) \rangle= \sum_k d_k|\varphi_k \rangle \langle \varphi_k|m;\varsigma\rangle
\label{psd}
\end{equation}
Because the choice of the basis $\{|\varphi_k\rangle\}$ is arbitrary, we may take $|\varphi_k\rangle =|k;\varsigma\rangle$ and hence obtain
\begin{equation}
|\psi_m(\varsigma)\rangle = d_m  |m;\varsigma\rangle
\label{psm}
\end{equation}
where the normalization of $|\psi_m(\varsigma) \rangle$ implies that $d_m$ is a mere phase factor, and hence can be put as $d_m=1$. 
\section{Jarzynski equality}\label{A3}
The Jarzynski equality (\ref{JE}) may be expressed in terms of the characteristic function (\ref{Gw}) as
\begin{equation}
G(i \beta) = Z(\lambda(\tau))/Z(\lambda(0))\:.
\label{Gib}
\end{equation}
In combination with (\ref{G}), (\ref{Q}), (\ref{R}) and the spectral representation of the initial Hamiltonian this condition becomes
\begin{equation}
\sum_m c_m p_m(\tau) =1
\label{cp}
\end{equation}
where 
\begin{equation}
p_m(\tau) = e^{-\beta \e_m(\tau)}/Z(\lambda(\tau))
\label{pm}
\end{equation}
denotes the equilibrium probability of the energy state $e_m(\tau)$. The coefficients $c_m$ are defined by
\begin{equation}
c_m= \sum_{n,k} e^{\beta(e_n(0) -e_k(0))} p(m,n|k)\:, \label{cmdefn}
\label{cm}
\end{equation}
where  
\begin{equation}
p(m,n|k) = \Tr U^\dagger(\Lambda) M^\dagger_m(\tau) M_m(\tau)U(\Lambda) M_n(0) \Pi_k(0) M_n^{\dagger}(0)
\label{pmnk}
\end{equation} 
determines the conditional probability to find $n$ in the initial and $m$ in the final measurement when the system was prepared in the eigenstate $|k;0\rangle$.

For universal measurements, the coefficients $c_m$ are independent of the final energies $e_m(\tau)$. Therefore, the condition (\ref{cp}) implies that $c_m=1$ for all $m$. Using (\ref{cmdefn}) this leads to
\begin{equation}
\sum_n p(m,n|n) + \sum_{n,k \atop n \neq k} e^{\beta(e_n(0) - e_k(0))} p(m,n|k) =1.
\label{spmnk0}
\end{equation} 
Due to the non-negativity of the conditional probabilities $p(m,n|k)$ and their independence of temperature it follows that
\begin{equation}
p(m,n|k) = 0 \quad \mbox{for}\; k\neq n\:.
\label{pmnk0}
\end{equation}
This implies 
\begin{equation}
p_0(n|k) = \sum_m p(m,n|k) =0  \quad \mbox{for}\; k\neq n\:,
\label{pnk0}
\end{equation}
and consequently, that the first measurement must be error-free.
Using the form of an error-free measurement, 
\begin{equation}
M_n(0) = |\psi_n(0) \rangle \langle n; 0|\:,
\end{equation}
the condition (\ref{spmnk0}) in combination with (\ref{pmnk}) becomes
\begin{equation}
\sum_n\langle \psi_n(0) |  U^\dagger(\Lambda) M^\dagger_m(\tau) M_m(\tau) U(\Lambda)  |\psi_n(0) \rangle  = 1\:.
\end{equation}
By means of the spectral representation of the propagator $U(\Lambda)= \sum_k e^{i \phi_k} |\varphi_k \rangle \langle \varphi_k |$ we obtain
\begin{equation}
\sum_{k,l} e^{-i(\phi_k -\phi_l)} \sum_n \langle \psi_n(0) |\varphi_k \rangle \langle \varphi_k |M^\dagger_m(\tau) M_m(\tau) |\varphi_l \rangle \langle \varphi_l|\psi_n(0) \rangle =1
\end{equation}
Because the phases $\phi_k$ are arbitrary in this double sum over $k$ and $l$ the diagonal terms must sum up to unity, while the non-diagonal elements must vanish individually. Hence, we obtain
\begin{eqnarray}
\sum_{n,k} |\langle \psi_n(0) | \varphi_k \rangle|^2 \langle \varphi_k |M^\dagger_m(\tau) M_m(\tau) | \varphi_k \rangle &= 1\:, \label{Ckk}\\
\langle \varphi_k |M^\dagger_m(\tau) M_m(\tau) |\varphi_l \rangle \sum_n \langle \psi_n(0) |\varphi_k \rangle \langle \varphi_l| \psi_n(0) \rangle &=0,\quad k\neq l\:. \label{Ckl}
\end{eqnarray}
In the second equation one can divide by the matrix element of $M^\dagger_m(\tau)M_m(\tau)$ because the basis $\{\varphi_k\}$ is arbitrary. 
This leads to
\begin{equation}
 \sum_n \langle \psi_n(0) |\varphi_k \rangle \langle \varphi_l| \psi_n(0) \rangle =0\:.
\label{psiphi0}
\end{equation}

Only for an operator proportional to unity, the non-diagonal elements vanish in any basis. Hence, we find
\begin{equation}
\sum_n |\psi_n(0) \rangle \langle \psi_n(0)| = c \mathbbm{1}\:.
\label{psi1}
\end{equation}
If the post-measurement states were not orthogonal to each other one could find an orthonormal basis with respect to which the left hand side of (\ref{psi1}) would have non-diagonal elements in contrast to the right hand side. Hence,  
the post-measurement states ${\psi_n}$ must represent a complete orthonormal basis with $c=1$.

Finally, from (\ref{Ckk}) in combination with the completeness relation, we obtain
\begin{equation}
\Tr M^\dagger_m(\tau) M_m(\tau) =1\:.
\end{equation} 
This completes the proof that the Jarzynski equality implies for universal measurements that (i) the initial measurements must be error-free, (ii)  the corresponding post-measurement states must form an orthonormal complete basis, and (iii) the effects of the second measurements must have unit trace. The sufficiency of these coefficients can be easily seen by inspection.


\section*{References} 
\begin{thebibliography}{99}       
\bibitem{J} Jarzynski C 1997 {\it Phys. Rev. Lett.} {\bf 78} 2690 
\bibitem{C} Crooks GE 1999 {\it Phys. Rev. E} {\bf 60} 2721 
\bibitem{JR} Jarzynski C 2011 {\it Ann. Rev. Cond. Mat. Phys.} {\bf 2} 329 
\bibitem{EHM} Esposito M, Harbola U and Mukamel S 2009 {\it Rev. Mod. Phys.} {\bf 81} 1665 
\bibitem{CHT} Campisi M, H\"anggi P and Talkner P 2011 {\it Rev. Mod. Phys.} {\bf 83} 771 
\bibitem{S} Seifert U 2012 {\it Rep. Prog. Phys.} {\bf 75} 126001 
\bibitem{Ciliberto} Douarche F, Ciliberto S, Petrosyan A and Rabiosi I 2005 {\it Europhys. Lett.} {\bf 70} 593 
\bibitem{BLR} Bustamante C, Liphardt J and Ritort F 2005 {\it Phys. Today} {\bf 58}(7) 43 
\bibitem{HSDL} Huber G, Schmidt-Kaler F, Deffner S and Lutz E 2008 {\it Phys. Rev. Lett.} {\bf 101} 070403 
\bibitem{K}  Kurchan J, arXiv:cond-mat/0007360
\bibitem{Tas} Tasaki H, arXiv:cond-mat/0009244
\bibitem{TLH} Talkner P, Lutz E and H\"anggi P 2007 {\it Phys. Rev. E} {\bf 75} 050102 
\bibitem{vN} von Neumann J 1996 {\it Mathematical Foundations of Quantum Mechanics}  (Princeton University Press, Princeton)
\bibitem{NC} Nielsen MA and Chuang IL 2000 {\it Quantum Computation and Quantum Information} (Cambridge University Press, Cambridge)
\bibitem{DCHFGV} Dorner R, Clark SR, Heaney L, Fazio R, Goold J and Vedral V 2013 {\it Phys. Rev. Lett.} {\bf 110} 230601 
\bibitem{MDP} Mazzola L, De Chiara G and  Paternostro M 2013  {\it Phys. Rev. Lett.} {\bf 110} 230602 
\bibitem{BSMASOGDPS} Batalh{\~a}o T, Souza AM,  Mazzola L,  Auccaise R,  Sarthour RS, Oliveira IS, Goold J, De Chiara G, Paternostro M and  Serra RM,  arXiv:1308.3241
\bibitem{CBKZH} Campisi M, Blattman R, Kohler S, Zueco D and H\"anggi P, arXiv:1307.2371
\bibitem{WM} Wiseman HM and  Milburn GJ 2010 {\it Quantum Measurement and Control} (Cambridge University Press, Cambridge)
\bibitem{LL} Landau LD and Lifshitz EM 1958 {\it Quantum Mechanics}, \S 7  (Pergamon Press, New York)
\bibitem{AG} Andrieux D and Gaspard P 2008 {\it Phys. Rev. Lett.} {\bf 100} 230404 
\bibitem{TMYH} Talkner P, Morillo M, Yi J and  H\"anggi P 2013 {\it New J. Phys.} {\bf 15} 095001 





\end{thebibliography}
\end{document}